**Approaches and principles of the *Meta-Structures Project*: the mesoscopic dynamics
-Notes for software and models designers-**

*Gianfranco Minati*
*Italian Systems Society, Via Pellegrino Rossi 42*
*Milan, I-20161, Italy*

**Abstract**-In this paper we specify research assumptions introduced and used in "The meta-structure project" to be considered both to *model* and *design* collective behaviours as given by coherent mesoscopic dynamics. We specify concepts of interaction and dynamics both in homogeneous and non-homogeneous cases, i.e., fixed and variable structures. We specify consequent modelling and design approaches. We also present and discuss approaches to measure the level of complexity allowed or detected in collective behaviours intended as coherent mesoscopic dynamics. We mention the possibility to model properties of domains of coherence, for instance, populations of oscillators and molecules, by using meta-structural properties to be eventually considered also for quantum coherent domains.
Keywords: Coherence, Collective, Dynamics, Homogeneous, Interaction, Mesoscopic, Meta-structural, Property.

**Introduction**

General and theoretical assumptions of the meta-structures project [Meta-structures Project, 2009] have been introduced in [Minati, 2008; 2009a; 2009b; 2011; Licata and Minati, 2010; Minati and Licata, 2011]. Within this conceptual framework, assumed known by the reader, in the first part of the paper we specify a) the concept of interaction, distinguishing between homogeneous and non-homogeneous cases, b) dynamics in the homogeneous and non-homogeneous cases, c) the conceptual case of populations of oscillators and d) the logical power of the approach related to computational emergence.

In the second part, we specify and discuss possible conceptual approaches to *design* and *model* mesoscopic dynamics. We consider the conceptual framework of spatial collective behaviours such as flocks, swarms, and traffic where metric and topological assumptions are possible. This is different, for instance, from collective behaviours of industrial districts dealing with economic aspects and collective energetic behaviours of towns [Minati and Collen, 2009]. We will specify rules of interaction, degrees of freedom and thresholds, and model dynamics as given by sequences of rules and process of selection among different states all possible by respecting the degrees of freedom and rules of interaction. We briefly specify the different degrees of complexity possible in this approach to mesoscopic dynamics.

In the Appendix we discuss the problem of fluctuations and coherence dominions to be eventually modelled by using meta-structural properties.

**1. Some theoretical assumptions of the approach**

The following theoretical assumptions are considered in a software engineering conceptual framework in order to specify and carry out simulation software to generate collective behaviours based on approaches and principles considered in the meta-structures project.
However the project's purpose is to consider meta-structures as a way to model generic collective behaviours and to allow influence on them in a non-invasive, non-destructive way making possible to orient, induce and vary processes of collective behaviours while maintaining their acquired properties. Other approaches focus on modelling emergence of meta-structures, intended as structures of structures [Johnson and Iravani, 2007; Pessa, 2011]

Approaches and assumptions are presented from an engineering point of view having, on one side, the eventual and potential, albeit very difficult, finality to implement simulation software and, on the other side, the aim to specify and exemplify the principles and approaches considered in the meta-structural analysis and modelling emergence of collective behaviours intended as conceptually generated by applying such assumptions and approaches. The meta-structure approach to model collective behaviour is *ideal* since it is based on a top-down structure considered conceptually valid to cover the widest possible spectrum of collective phenomena. In ideal models this makes it possible to allow for the deduction of particular consequences and for forecasting when suitable mathematical tools and approaches are available. However, when an ideal model of emergence produces a result, one can be sure that this is not a consequence of some *ad hoc* assumption, but derives from its basic principles and approaches. Examples of other ideal models are given by Noise-induced phase transitions, Spontaneous Symmetry Breaking in Quantum Field Theory, and, eventually, Multiple Systems [Minati and Pessa, 2006].

**1.2 Interactions: the general framework**

A collective behaviour is considered here as given by coherent sequences of spatial configurations adopted by agents interacting through corresponding different structures over time, as in the meta-structure approach [Minati and Licata, 2012].

The initial state at $t_0$ coincides with a specific configuration, for instance, of a population of *n* collectively interacting agents establishing collective behaviour, i.e., acquiring emergent properties such as behaviour and pattern. In this Section we will consider the conceptual framework and approaches assumed in the meta-structure project. Technical aspects are discussed in Section 2.

**1.2.1 Homogeneity**

In classical approaches used to model and simulate phenomena of collective behaviour, both by using systems of differential equations and, for instance, Cellular Automata and Neural Networks, coherence is assumed be established *instantly* and kept by the same, simultaneous and global, continuous interactions.

Coherence is given by a coherent sequence of configurational states reached by accomplished interactions.

The assumption is that all interactions always occur identically, by applying the same rules and eventual parametrical changes at each instant.

In the homogeneity case, the conceptual framework is that all changes occur in a *field of same rules* eventually applied with parametrical, regular changes suitable to avoid violation of degrees of freedom or to deal with external environmental constraints, such as obstacles or the coming of a predator.

*Dynamics is given by application of the same rules of interaction to the previous configuration*, i.e., to agents possessing different metrical and topological states.

**1.2.2 Dynamics of the homogeneity case**

The same rules are identically and simultaneously applied to all elements.
All elements always respect all degrees of freedom.
Interactions occur when *each* element assumes, i.e., *choices* among all the possible ones a new state at the time $t_{i+1}$ depending on the states of *all* other elements at time $t_i$. Each element at the time $t_i$ possesses a space of states possible at the time $t_{i+1}$.
The process of interaction takes place simultaneously for all elements.
Coherence is conceptually given by the identification of a subspace of the space of all states possible for each element, such that states possible for each agent at time $t_{i+1}$ will always respect degrees of freedom and rules of interaction.
The process of choice of a state for each agent in such a subspace represents the dynamics of the collective behaviour, i.e., the shift from the state of the collective behaviour from the time $t_i$ to the $t_i$ $_{+1}$. The process of choice consists of the *same* choice function based on random parametrical variations.

*All* elements vary their state at the time $t_{i+1}$ depending on the state of *all* other elements at time $t_i$.

In this case coherence is instantaneous.

The space state in the discrete case may be modelled by using, for instance, both systems of differential equations and Cellular Automata or Neural Networks. Different approaches may be assumed for the choice function. A particular case occurs when interactions are not instantaneous, but *propagate* in any possible way, for instance by sequential, discretised diffusion along time. *In this later case the system keeps memory, as strategy, of any eventual number of the previous steps in order to allow homogeneity and, even, apply context-sensitive sequences of interaction, for instance depending on density and obstacles.*

### 1.3.1 Non-homogeneity

It is possible to introduce variations to the previous conceptual framework, such as the following:

- The application of rules of interaction to allow the shift from a configuration to another may not necessarily identify for each agent a *unique* new state, but a set of new possible ones. Dynamics is given in this case by applying a *choice function* to the set of all possible states, all equivalent since all satisfying rules of interactions and the degrees of freedom. Examples of choice functions are given by a) random functions, b) considering the minimum energy and then selecting the nearest position, c) considering previous selections, avoiding, for instance, immediate repetition of criteria or repeating the criterion previously used for the closest one.
- It is possible to assume that at each step the mix of rules of interaction applied to specific cases changes. For instance, it is possible to compute possible positions of an agent at the following instant by complying with different percentages of usage of the degrees of freedom, different thresholds and parameters setting values of minimum and maximum distances, and direction. The choice function is then applied on these possible different states computed.
- It is also possible to assume the application of different approaches in a non-homogenous and variable way to different and variable clusters of agents [Mikhailov and Calenbuhr, 2002; Minati and Licata, 2012]. It is possible to ideally consider any distributions of approaches and clustering. We can cite as an example combinations of criteria sequencing rules and considering topological criteria, non-local repeatability starting from different initial states, global repeatability and statistical distribution. *A notable case takes place when the same mix of rules is maintained for a variable number of sequences for the same agent*. The situations described above are represented by the variation of the mesoscopic general vector introduced in [Minati and Licata, 2012].

Transient states are given by the *break of the interactional regime*, i.e., by the shift from a regime of interactions to another one unlike the homogeneous case. Examples are given by cases of non-homogeneity listed above and by sequences of different interactions.

*The number and frequency of breaks of the interactional regime states the level of complexity of the process of non-radical*, see 1.3, *emergence of the collective behaviour.*

In a transient state there is simultaneous validity of two or more interactional regimes eventually at different scales. In the homogeneous case transient states, sources of the level of complexity, do not occur.

In the homogeneous case transient states, *in principle*, do not exist.

*Dynamics is given by application of the choice function and different rules of interaction to the previous configuration.*

In the meta-structure project the configurational states, i.e., sequences of configurations, assumed both in the case of homogeneity and non-homogeneity are represented by values acquired by mesoscopic variables and their eventual coherence, i.e., emergence of acquired properties kept along time, is represented by suitable meta-structural properties [Minati and Licata, 2012].

**1.3.2 Dynamics of the non-homogeneous case**

Rules are not the same per instant, per agent and by considering sequences.

In the non-homogeneous case not only interactions change, but they do not apply simultaneously.

Different possibilities may be considered such as:
- Interactional regimes may apply to different clusters at different times;
- The sequence of interactional regimes and relative clusters may be due to any choice function responsible to maintain coherence;
- Same clusters at subsequent times may do not apply to any interactional regime, but just linearly *continue* the previous trajectories.
- The process of choice may consist of *different* choice functions, rather than choices made by the same function.

Another case occurs when some elements maintain or change their behaviour respecting degrees of freedom *only*, such as for leaders activating gregarious behaviour. In this case the other elements compute their behaviour, based on rules of interaction, in order to keep coherence, i.e., the general respect of degree of freedom *and* rules of interaction.

**1.4 The case of populations of oscillators**

It is possible to exemplify the homogeneity and the non-homogeneity cases by considering populations of oscillatory.

The case related to homogeneity may be considered consisting of:
- Populations of oscillatory in phase**;**
- Populations of clusters of oscillatory in different phases. The difference of phases between clusters is constant or regular, e.g., periodic.

The case related to non-homogeneity may be considered consisting of:
- Populations of oscillators dynamically clustered corresponding to phase variations. The differences between phases are not regular, e.g., random or established by a choice function among the ones possible depending on rules of interaction;
- Populations of oscillators where initial phase conditions of cycles, i.e., interactional regimes, are computed, for instance, in function of the phases possessed by adjacent or oscillators at previous times.

A transient state between cycles is assumed to occur when, for instance, there is change of choice function.

**1.5 Logical power of the approach. Detecting non-radical emergence.**

In the case of homogeneity there are conceptually *identical* simulations from *identical* initial conditions. It is eventually possible to have *different* simulated behaviours from *identical* initial conditions by introducing random parametric variations and *changes of usage of degrees of freedom* [Minati and Licata, 2012] always complying with thresholds and usage of degrees of freedom.

In the case of complete sameness of conditions, simulations will be identical.

In the case of non-homogeneity even from identical initial conditions, different simulations will be different since the cases listed at 1.2.2 considered above allow establishment of different sequences of structure, configurations and usages of degrees of freedom.

However, it is important to point out that the approach considered in the meta-structures project allows detections and modelling of non-radical emergence, such as phenomenological emergence. As a matter of fact, variables considered in the project, both in the case of maintenance and break of the interactional regime, are the *same* or considered at different micro-, macro-, and mesoscopic scales.

There is a *gestaltic continuity* between the three scales even if clustering may occur in different ways, for instance by minimizing the energy spent by the observer in the neuronal phase space [Edelman and Tononi, 2000], and considering different interactions, their mixes and sequences.

The approach is conceptually also usable to model processes a) where to study eventual emergence without detection of acquisition of new properties at the scale of variables and scale-free, and b) assumed established by sequences of suitable variables and structures to be identified and possessing meta-structural properties, such as images and sounds [Minati and Licata, 2012].

Models based on meta-structures are suitable both for the homogeneity and non-homogeneity cases.

The approach, instead, is not suitable to deal with *radical emergence*. Radical emergence occurs, for instance, when the continuity mentioned above is broken at any scale. It is not possible to linearly or non-linearly relate new variables, models and properties to ones used before emergence. This is typically the case for quantum phenomena where state variable values are not accessible at any moment and in the same way due to uncertainty principles. Instead of considering the set of state variables, it is possible to study the system state characterised by the wave-function or state-vector $\Psi$ considered in quantum physics [Licata and Sakaji, 2008]. Examples are given by emergence of consciousness and of phenomena where there is emergence of eventual *same* properties previously acquired through processes modelled by non-radical emergence, but possessing completely different and non-reducible characteristics such as for superconductivity, superfluidity and change of resistance in conductors at the microscopic scale of 1000 nm.

It conceptually corresponds, in the approach considered here, to change mesoscopic variables and not only structures and their sequences. This extreme case will be considered in future works.

**2. Modelling and Simulating. Mesoscopic and meta-structural dynamics**

In this Section we specify effective approaches to model and simulate the theoretical assumptions introduced above suitable for software engineering. In Sections 2.1.1 we specify *effective* rules of interaction and in 2.1.2 degrees of freedom assumed to allow simulated collective behaviours. In Section 2.2 we specify the mesoscopic modelling approach, the mesoscopic dynamics and the detection of complexity. We will also mention the theoretical related problem of decidability.

**2.1 Collective interactions**

As introduced above, with regards to the homogeneous case, we will consider the subspace of possible states given by the *effective* application of degrees of freedom and rules of interaction, only conceptually introduced above. Within this subspace, the function of choice allows the effective dynamics of the collective behaviour.

With regard to the non-homogeneous case, the subspace considered above is given by a dynamical set of subspaces given by clustering, changes of the choice functions in charge to select possible points and sequences of interactional regimes. Dynamics is given by sequences of changes computed by following rules of interaction, clustering, degrees of freedom and choice functions.

**2.1.1 Rules of interaction**

Starting from any configuration, the application of one or more rules of interaction to one or more elements changes the configuration with respect to degrees of freedom. We underline that it is not reduced to *verify* that the current configuration respects the rules of interactions. Rules of interactions are always applied to compute, *activate* a new configuration.

Rules of interaction specify the change of, for instance, position and direction of an element depending on other one. When an element changes property, all others must change depending on the rules of interaction.

We will consider the follow examples based on classical approaches introduced in the literature regarding collective behaviours in *3D*, e.g., swarms and flocks [Reynolds, 1987; 1999]. Rules of interaction *simultaneous* applied during the interaction may be:

*Alignment*. Compute the average direction of the displacement assumed by adjacent elements at the time $t_i$ and assumption of the new position at the time $t_{i+1}$, *close*, i.e., different less than a threshold value $ATDir_{\Delta i+1}$, to this direction. This ideal approach applies in the homogeneous case and in the non-homogeneous case to sequences of clusters.

<div style="text-align:center">and</div>

*Cohesion*. Compute the average position assumed by adjacent elements at the time $t_i$ and assumption of the new position at the time $t_{i+1}$, *close*, i.e., different less than a threshold value $ATPos_{\Delta i+1}$, to this position. This ideal approach applies in the homogeneous case and in the non-homogeneous case to sequences of clusters.

Thresholds $ATDir_{\Delta i+1}$ and $ATPos_{\Delta i+1}$ may be:
- different per instant, $ATDir_{\Delta i+1} \neq ATPos_{\Delta i+1}$;
- different along time, $ATDir_{\Delta i} \neq ATDir_{\Delta i+1}$ and $ATPos_{\Delta i} \neq ATPos_{\Delta i+1}$;
- different for clusters both at the same time and along time.

<div style="text-align:center">and</div>

*Separation*. Compute the new position in such a way to guarantee density of adjacent elements less than a suitable global threshold value, fixed, variable along time, or variable per clusters and eventually variable along time.

<div style="text-align:center">and</div>

*Topological*. Compute the new direction, position, speed and altitude in such a way to guarantee a topological property of single elements such as:
- Belonging to the geometrical surface or the centre;
- Having a specific cognitive topological distance from one of the agents belonging to the geometrical surface or centre;
- Be at the *cognitive topological centre* of the collective behaviour, i.e., all cognitive topological distances between the agent under study and all the agents belonging to the geometrical surface are equal. This body may be *virtual* and be considered as a *cognitive attractor* for the collective behaviour. Its trajectory may represent the trajectory of the collective behaviour [Ballerini *et al.*, 2008].

**2.1.2 Degrees of freedom**

Degrees of freedom do not activate any change, they limit possible changes. Examples are:
- Distances, directions, speeds, and altitudes possessed by single element *always change* over time.
- Maximum and minimum limits are established.
- These degrees of freedom may apply in different ways along time, such as regarding clusters of agents and partially.

## 2.2 Modelling and representing mesoscopic dynamics

Coherence of sequences of configurations establishing collective behaviour is considered in the meta-structural approach as represented and given by suitable meta-structural properties. In this view:
- Suitable mesoscopic variables *transversally intercept* and represent values assumed by aggregates of microscopic variables. Values of mesoscopic variables then represent the effective applications of rules of interaction.
- Suitable properties of sets of such values represent coherence of sequences of configurations, i.e., the collective behaviour.

While coherence of single systems is assumed given by the single, stable corresponding structure, i.e., rules of interaction, constraints and suitable parameters imposed by the designer or assumed by the modeller, making elements to interact always in the *same* way, coherence of collective behaviour is considered due to coherence of multiple and superimposed, i.e., *simultaneous* like in Multiple Systems, structures making elements to interact in different and multiple ways. Multiple structures are considered suitably represented by mesoscopic variables specifying effective applications of rules of interaction, such as, at a suitable threshold:

a) $Mx(t_i)$ number of elements having the maximum distance at a given point in time;
b) $Mn(t_i)$ number of elements having the minimum distance at a given point in time;
c) $N_1(t_i)$ number of elements having the *same* distance from the nearest neighbour at a given point in time;
d) $N_2(t_i)$ number of elements having the *same* speed at a given point in time;
e) $N_3(t_i)$ number of elements having the *same* direction at a given point in time;
f) $N_4(t_i)$ number of elements having the *same* altitude at a given point in time;
g) $N_5(t_i)$ number of elements having the *same* topological position at a given point in time.

It is important to stress that a mesoscopic state variable at time $t_i$ may assume different values representing the number of elements having the *same* value (*equal* when *within a range of values or thresholds*) such as distance, altitude and speed. Furthermore *n*-elements constituting a mesoscopic state variable at instant $t_i$ may be clustered into groups having the *same* values. By considering distance, it is possible at time $t_i$ to have:
- $n_1$ elements at distance $d_1$,
- $n_2$ are at distance $d_2$, etc.

The value assumed by the mesoscopic state variable can be:
- $n_1 + n_2 + \ldots + n_s$ may be $> n$ (same elements constitute different clusters),
- $< n$ (not all elements constitute different clusters) or $= n$ (each element belongs to one cluster only).

Dealing now with the dynamics of a collective behaviour, i.e., the coherent sequence of configurations, it is possible to consider almost four exemplificative phases.

1) All the agents simultaneously possess all the same mesoscopic properties *constant* along time. There is simultaneously respect of the degrees of freedom *and* parametrical values defining mesoscopic variables are *constant*, e.g., it is considered that the *same* distance and altitude may *insignificantly* only change within the threshold assumed.
   For any agent *1,k* and for $\forall$ mesoscopic property $m$, $V_{k,m}(t_i) = [1,1,1, \ldots, 1]$.

2) All the agents simultaneously possess all the same mesoscopic properties *constant* per instant, but *variable* along time. There is simultaneously respect of the degrees of freedom *and* parametrical values defining mesoscopic variables are *constant* per instant, but *variable* along time. e.g., it is considered that the *same* distance and altitude may *insignificantly* only change within the threshold assumed *per instant*, while they can significantly change along time.
   For any agent *1,k* and for $\forall$ mesoscopic property $m_i$, $V_{k,m}(t_i) = [1,1,1, \ldots, 1]$, where $m(t_i) \neq m(t_{i+1})$.

3) Not all the agents non-simultaneously possess not all the same mesoscopic properties *constant* along time. There is simultaneously respect of the degrees of freedom *and* parametrical values defining mesoscopic variables are *constant*, e.g., it is considered that the *same* distance and altitude may *insignificantly* only change within the threshold assumed. For any agent *1,k* and for $\forall$ mesoscopic property *m*, it will occur per instant different configurational varieties of the vector $V_{k,m}(t_i) = [e_{k,1}(t_i), e_{k,2}(t_i), ..., e_{k,m}(t_i)]$ such as:

   $V_{1,m}(t_i) = [1,0,0, ..., 0]$
   $V_{2,m}(t_i) = [0,1,0, ..., 1]$
   $V_{3,m}(t_i) = [0,1,1, ..., 0]$
   $V_{4,m}(t_i) = [1,0,1, ..., 1]$
   $V_{5,m}(t_i) = [0,0,0, ..., 1]$
   ...
   $V_{k,m}(t_i) = [0,1,0, ..., 1]$.

4) Not all the agents non-simultaneously posses not all the same mesoscopic properties *constant* per instant, but *variable* along time. There is simultaneously respect of the degrees of freedom *and* parametrical values defining mesoscopic variables are *constant* per instant, but *variable* along time. e.g., it is considered that the *same* distance and altitude may *insignificantly* only change within the threshold assumed *per instant*, while they can significantly change along time.

   For any agent *1,k* and for $\forall$ mesoscopic property $m(t_i) \neq m(t_{i+1})$, it will occur per instant different configurational varieties of the vector $V_{k,mi}(t_i) = [e_{k,1}(t_i), e_{k,2}(t_i), ..., e_{k,m}(t_i)]$ such as:

   $V_{1,mi}(t_i) = [1,0,0, ..., 0]$
   $V_{2,mi}(t_i) = [0,1,0, ..., 1]$
   $V_{3,mi}(t_i) = [0,1,1, ..., 0]$
   $V_{4,mi}(t_i) = [1,0,1, ..., 1]$
   $V_{5,mi}(t_i) = [0,0,0, ..., 1]$
   ...
   $V_{k,mi}(t_i) = [0,1,0, ..., 1]$.

We mention an interesting future research issue relating to the study of the conceptual, possible correspondence between the four cases listed above and the classical four classes of complexity introduced by Stephen Wolfram for Cellular Automata [see, for instance, Wolfram 2002].

| Mesoscopic Dynamics | | | |
|---|---|---|---|
| Case 1 | Case 2 | Case 3 | Case 4 |
| *All* the agents *simultaneously* possess *all* the same mesoscopic properties *constant* along time (only *insignificant* changes within the threshold assumed) | *All* the agents *simultaneously* possess *all* the same mesoscopic properties *constant* per instant, but *variable* along time (mesoscopic variables are *constant* per instant, but *variable* along time) | Not all the agents non-simultaneously possess not all the same mesoscopic properties *constant* along time (only *insignificant* changes within the threshold assumed) | Not all the agents non-simultaneously possess not all the same mesoscopic properties *constant* per instant, but *variable* along time (mesoscopic variables are *constant* per instant, but *variable* along time) |
| *Mesoscopic structure fixed* | *Changes of the same mesoscopic structure* | *Multiple and superimposed variations of the same mesoscopic structures* | *Multiple and superimposed variations of the mesoscopic structures* |
| Trivial meta-structural properties. *Collective behaviours structurally 'fixed'* | Trivial meta-structural properties. *Collective behaviours structurally at low variability* | Non-trivial meta-structural properties. *Collective behaviours structurally variable* | Non-trivial meta-structural properties. *Collective behaviours structurally at high variability* |

→ Direction representing increasing of complexity due to increasing of structural change

*Tab. 1- Mesoscopic Dynamics*

Meta-structural properties, i.e., mathematical properties of the sets of values assumed by mesoscopic variables over time, *represent coherence* of the mesoscopic dynamics → which represents coherence of sequences of multiple and superimposed structures → which produces microscopic coherence.   We also mention two interesting research issues:
1. Mesoscopic degrees of freedom
    It is possible to introduce *mesoscopic degrees of freedom* regarding, for instance, minimum and maximum of values assumed by mesoscopic variables; general limitations to groups of values assumed by mesoscopic variables, e.g., their sum or weighted combinations, and to ratios among them.
2. Decidability
    It is possible, by using mesoscopic variables and meta-structural properties, to make collective behaviours *decidable*? That is, is it possible to decide in a finite number of computational steps if a population of interacting agents, at a suitable scalarity, establishes a collective behaviour, i.e., possesses suitable meta-structural properties?

**Conclusions**

In this paper we presented, having modellers and designers in mind, theoretical approaches based on considering collective behaviours as coherent mesoscopic dynamics. The paper contains specifications useful for researchers interested in considering meta-structures as conceptual framework to model, and eventually consider as such, phenomena of collective behaviour. The same specifications are useful to design collective behaviours by both a) directly and explicitly *prescribing* mesoscopic coherence and b) *inducing* emergence of collective behaviour when prescribing meta-structural properties. We mention further possible areas of research such as mesoscopic degrees of freedom *when design*, decidability *when detecting and modelling*, and coherence domains eventually modelled by meta-structural properties.

**Appendix: Fluctuations and Coherence Domains**

Differently from the concept of mesoscopic variable used here in the meta-structure project and intended as *aggregate of microscopic variables*, in physics a generic aspect or property is termed mesoscopic, and the related observation scale is mesoscopic, when it is not possible to ignore the role of *fluctuations* generating measurable physical effects, not necessary small, but often very remarkable. In this case mesoscopic variables are physical quantities describing the state of a material or dynamics of a process taking place at the mesoscopic scale, i.e. between molecular size and micrometer range, where quantum mechanical phenomena start to be revealed.

We consider stochastic instabilities of phenomena as fluctuations describable by using different approaches. In particular, quantum fluctuations are temporary changes of the energetic state assumed by the vacuum and occurring with respect of the Heisenberg uncertainty principle.

This quantum uncertainty allows the appearing in the vacuum of small quantities of energy if they disappear in a very short time, i.e., fluctuations.

As a consequence, wherever introducing a mesoscopic variable, its evolution will necessarily also be influenced by fluctuations as an unavoidable component.

Collective behaviours such as flocks and swarms are macroscopic entities at a suitable scale, for instance at a scalarity larger than the average distance between components. At a smaller scalarity, for instance significantly smaller than the average distance between components, we could notice that mutual distances between elements are affected by fluctuations and their evolution is stochastic.

An interesting corresponding research issue relates to consider meta-structures, i.e., coherent sequences of structures, as coherence quantum domains when dealing with quantum mesoscopic variables. In this case the research issue relates to the possibility to consider meta-structural properties or domains of coherence quantum domains [Sewell, 2002].